\documentclass[aps,prc,twocolumn,showpacs,superscriptaddress,reprint]{revtex4-1}
\usepackage{amsmath}
\usepackage{epsfig}  
\usepackage{color}
\usepackage{endnotes}
\usepackage{natbib}
\usepackage[colorlinks,breaklinks]{hyperref}
\usepackage{nicefrac}
\usepackage{bm}% bold math
\usepackage{dcolumn}
\usepackage{graphics}% Include figure files
\usepackage{amsmath}
\usepackage{amsfonts}
\usepackage{amssymb}
\usepackage{color}
\newcommand{\bra}{\langle}
\newcommand{\ket}{\rangle}

\newcommand{\bs}[1]{\ensuremath{\boldsymbol{#1}}}

\newcommand{\be}{\begin{equation}}
\newcommand{\ee}{\end{equation}}
\newcommand{\bea}{\begin{align}}
\newcommand{\eea}{\end{align}}
\newcommand{\beqa}{\begin{eqnarray}}
\newcommand{\eeqa}{\end{eqnarray}}
\newcommand{\nablavec}{\ensuremath{\boldsymbol{\nabla}}}
\newcommand{\rvec}{\ensuremath{\boldsymbol{r}}}
\newcommand{\xvec}{\ensuremath{\boldsymbol{x}}}
\newcommand{\yvec}{\ensuremath{\boldsymbol{y}}}
\newcommand{\zvec}{\ensuremath{\boldsymbol{z}}}
\newcommand{\nvec}{\ensuremath{\boldsymbol{n}}}

\newcommand{\dvec}{\ensuremath{\boldsymbol{d}}}
\newcommand{\svec}{\ensuremath{\boldsymbol{s}}}
\newcommand{\tvec}{\ensuremath{\boldsymbol{t}}}

\begin{document}

\title{Extrapolating Lattice QCD Results using Effective Field Theory}

\date{\today}

\author{Moti Eliyahu}
\affiliation{The Racah Institute of Physics, The Hebrew University, 
Jerusalem 9190401, Israel}

\author{Betzalel Bazak}
\affiliation{The Racah Institute of Physics, The Hebrew University, 
Jerusalem 9190401, Israel}

\author{Nir Barnea}
\affiliation{The Racah Institute of Physics, The Hebrew University, 
Jerusalem 9190401, Israel}

\begin{abstract}
  Lattice simulations are the only viable way to obtain ab-initio Quantum
  Chromodynamics (QCD) predictions for low energy nuclear physics. These 
  calculations are done, however, in a finite box and therefore extrapolation is needed
  to get the free space results.
  Here we use nuclear Effective Field Theory (EFT), designed to
  provide a low energy description of QCD using baryonic degrees of freedom, to
  extrapolate the lattice results from finite to infinite volumes.
  To this end, we fit the EFT to the results calculated with nonphysical high quark 
  masses and solve it with the stochastic variational method in both finite and infinite volumes.
  Moreover, we perform similar EFT calculations of the physical point and 
  predict the finite-volume effects to be found in future Lattice QCD calculations
  for atomic nuclei with mass number $A\le4$.
\end{abstract}

\maketitle

%============
\section{Introduction}
%============
At low energies, characterizing the nuclear structure, Quantum Chromodynamics (QCD),
the fundamental theory of the strong interactions, is non-perturbative. The only feasible way 
to obtain \emph{ab-initio} QCD predictions for nuclear physics is through Lattice
simulations of QCD, dubbed LQCD \cite{Wil74}.

These calculations are done via numerical evaluation of path integrals on a discrete
space- and time-like lattice and summation over all possible paths. When the volume of the
lattice is taken to be infinitely large and its sites infinitesimally close to each other,
the continuum is recovered.

After years of development, LQCD simulations are fulfilling their promise of calculating
static and dynamical quantities with controlled approximations. Progress has been made to
a point where meson and single-baryon properties can be predicted quite accurately; see, e.g.,
\cite{Dur08,FodHoe12,Bor15}. 
However, the complexity and peculiar fine-tuning aspects in nuclear systems make this
fundamental approach significantly more difficult relative to the extraction of
single-baryon observables. For a recent reviews, see, e.g. \cite{LinMey15,Dav18}.

Currently, a few LQCD collaborations are studying multi-baryon systems, including
HAL QCD \cite{HAL11,HAL12},
PACS \cite{YamIshKur12,YamIshKur15},
NPLQCD \cite{NPLQCD,OrgParSav15,NPLQCD18},
CalLat \cite{CalLat17}, and 
the Mainz group \cite{Mainz19}. 
Most teams try to extract the nuclear binding energies directly from the lattice simulations.
The HAL QCD collaboration takes a different approach, 
trying to extract the nuclear interaction from the lattice
simulation, and then calculate observables using
standard nuclear physics techniques with the resulting nucleon-nucleon potential. 
At this point, HAL QCD results
are different from the results of the other groups.

A more common approach to study nuclear physics is based on Effective Field
Theories (EFTs). In nuclear EFTs, baryons and mesons replace the quarks and
gluons as the fundamental degrees of freedom. This framework provides a
practical theory to analyze nuclear physics while incorporating the essential
features of QCD.
For low energy aspects of nuclear physics, like the description of light nuclei,
even the mesons are not needed, and one is left with baryonic EFT, 
commonly referred to as pionless EFT, which will be employed here. 
This EFT is especially appropriate in a heavy pion mass world, where pion
dynamics are suppressed. 

The first application of EFT to multi-baryon LQCD faced the challenge of extending LQCD results 
to study the binding energies of larger $A>4$ nuclei. 
A pionless EFT was fitted to the LQCD outcome and then used to predict the ground-state
energies of $^5$He and $^6$Li \cite{BarConGaz15}, as well as $^{16}$O \cite{ConLovPed18}.  

Here we would like to use similar EFT to deal with another aspect of nuclear LQCD calculations.
LQCD calculations necessarily take place in finite volumes, thus affecting
their infrared properties. For two-body systems, it is fair to claim that the implications of
the finite volume on the spectrum are well understood through the L\"uscher formalism
\cite{Lus86,Lus91}.
The formalism pertinent for systems beyond the two-body
system has not yet reached this level of maturity,
while significant progress is achieved in recent years, mainly in the three-body system, see e.g., 
\cite{BriDav13,HanSha14,MeiRioRus15,KorLuu16,BriHanSha17,HamPanRus17,KonLee18}. 

The complexity of the problem calls for an alternative road map towards the determination
of infinite-volume quantities.
Such an approach could be constructing a nuclear EFT having the same boundary conditions as LQCD.
This way the EFT is built directly matching the LQCD results in a finite lattice,
and the extrapolation to the infinite lattice can be easily carried out through
the nuclear EFT.
Doing so, LQCD calculations may be performed with smaller lattice volumes, giving more
accurate results, leaving the extrapolation to be done by the EFT.

Here we use the NPLQCD results for pion mass of $m_\pi = 806$ MeV \cite{NPLQCD} to calibrate 
a leading order pionless EFT at finite box size and extrapolate the results toward the
free space limit. 
Moreover, we perform the inverse procedure for the case of physical pion mass, i.e. we fit our
EFT to the experimental results in the continuum and then predict the finite-box effect to  
be calculated in future LQCD calculations. 
Calculations are performed for bound nuclei with mass number $A \le 4$.

%=========
\section{Theory}
\label{sec:thoey}
%=========

As mentioned above, LQCD calculations are done in a finite volume, thus finite size 
effect should be corrected in order to extract the relevant physical quantities.

The common approach to do so is based on L\"uscher's work \cite{Lus86,Lus91}, which
solves the two-body problem in a large box.
This way one can get the first order correction to the free-space binding energy,
\be\label{Luscher}
E_B - E_L = -24\pi |A|^2 \frac{e^{-\kappa L}}{mL}+O\left(e^{-\sqrt 2 \kappa L}\right)
\ee
where
$E_B$ is the free space binding energy, $E_L$ is the binding energy on a lattice with size $L$,
$|A|^2 \sim 1$ is a normalization constant, and 
$\kappa=\sqrt{mE_B}$ is the binding momentum.
To utilize L\"uscher formula one has to calculate the binding energies for a few lattice sizes
and fit the results with Eq. \eqref{Luscher} to extract the free space parameters.
Similar method enables the extraction of free space scattering parameters from
finite box bound state calculations, avoiding the complication of dealing with continuum
states. 
Another method is to calculate in a single lattice several boosted states differ by their total momentum.
Using the asymptotic solution of boosted states in a box, the free space parameters can
be extracted \cite{BouKonLee11,DavSav11}. 
A twisted boundary conditions were also applied and shown to give better convergence \cite{BriDavLuu14}.

The three-body case can be solved in some simple cases. Based on that, several methods
to correct the finite lattice effects were developed  
\cite{BriDav13,HanSha14,MeiRioRus15,KorLuu16,BriHanSha17,HamPanRus17}. 
The generalization of these methods to larger systems is an open challenge.

Here we would like to take a different path, relevant to arbitrary
particle number, based on the construction of a relevant
EFT which can be solved in any lattice size.

Effective field theories are a powerful tool to study the low energy properties of a system
whenever a separation of scales exists between the energy scale of the process under
inspection and the typical scale of the underlying theory.

Weinberg \cite{Wei90} has formulated the idea that in order to calculate low-energy
observables of a given theory, it is sufficient to write down the most general Lagrangian,
whose form is only limited by general properties like analyticity, unitarity, and 
the symmetries of the theory under investigation. In the case of QCD, it is Lorentz symmetry,
parity, time reversal, and charge conjugation.
Chiral symmetry, which is an approximated symmetry for the physical $u$ and $d$ quarks, 
does not apply
in our case of heavy quarks. 
The fields used as degrees of freedom in this
effective Lagrangian should be those which are seen as asymptotic states in the regime one is
interested in.
For low-energy nuclear physics, the relevant degrees of freedom are 
the nucleons.

A general Lagrangian constructed this way contains an infinite number of terms. The key
ingredient to
resolve this obstacle is the scale separation mentioned above: Being only interested in
low-energy observables, one can assume that the terms in the Lagrangian are ordered by a
small parameter, which is the ratio of the energy scales involved. 

Pionless EFT is the resulting theory for baryon–baryon interactions. It does not contain explicit 
pions only contact interactions.
The process of establishing the order of terms in the EFT is called power counting. 
For pionless EFT it is well known that the naive power counting, based on counting powers of momentum, 
fails due to the emergence of Efimov physics \cite{Efimov70}. The three-body
contact term is to be promoted to leading order \cite{BHvK99}; see, however, \cite{BBB19}.

The relevant Lagrangian at leading order is therefore,
\be \begin{split}
\mathcal L = N^\dagger \left(i\partial_0 
+ \frac{\nablavec^2}{2m} \right) N
&- \frac{C_0}{2} (N^\dagger N)^2
- \frac{C_1}{2} (N^\dagger {\bf \sigma} N)^2 -  \\
&- \frac{D}{6} (N^\dagger N)^3 
\label{Leff} \end{split}
\ee
where $N$ is the nucleon field operator, and
$C_0$, $C_1$ and $D$ are the low-energy constants (LECs).
This Lagrangian can be supplemented with terms containing more fields and/or more derivatives, which are subdominant. Since in this work we
focus on the leading order, such terms will be neglected in the following. 

Contact interactions are singular and therefore regularization is needed; here
we use a Gaussian regulator $\tilde g(p)=\exp[-(p/\Lambda)^2]$
that suppresses momenta above an ultraviolet cutoff $\Lambda$.
Since the cutoff is not a physical quantity, the theory observables have to be
independent of it. This is achieved via renormalization, i.e. by fitting the values of 
the LECs 
$C_0 = C_0(\Lambda)$, $C_1 = C_1(\Lambda)$ and 
$D = D(\Lambda)$ to a chosen set of physical observables.

The leading order interaction in pionless EFT is to be iterated, which is equivalent 
to solving the non-relativistic Schr\"odinger equation with the Hamiltonian 
\be \label{H}
H = -\frac{1}{2m}\sum_i \nabla^2_i + 
\sum_{i<j} V_2(r_{ij}) + \sum_{i<j<k} V_3(r_{ij},r_{jk}).
\ee
Here 
\be\label{v2}
V_2(r_{ij}) = \left( C_0 + C_1 \sigma_i \cdot \sigma_j \right) g_\Lambda(r_{ij})
\ee
is the two-body interaction,
\be\label{v3}
V_3(r_{ij},r_{jk}) = D \sum_{cyc} g_\Lambda(r_{ij}) g_\Lambda(r_{jk})
\ee
is the three-body interaction, 
$g_\Lambda(r)=\frac{\Lambda^3}{8 \pi^{3/2}}\exp(-\Lambda^2r^2/4)$,
and $\sum_{cyc}$ stands for cyclic permutation of $\{i,j,k\}$.

Putting our system in a box with periodic boundary conditions, one has to solve
the eigenvalue problem 
\be
H_L \Psi_L = E_L \Psi_L
\ee
where the subscript $L$ denotes the lattice. On the lattice, the wavefunction $\Psi_L$ is to 
obey the periodic boundary conditions,
\be
\Psi_L(\rvec_1, \rvec_2, \dots) =  \Psi_L(\rvec_1+\nvec_1 L, \rvec_2+ \nvec_2 L, \dots)
\ee
for arbitrary integers trios $\{\nvec_1,\nvec_2,\ldots\}$.
The Hamiltonian $H_L$ is composed of the regular kinetic energy and the periodic 
potential $V_L$ given by
\be
    V_L(\rvec_1,\rvec_2,\ldots ) = \sum_{\nvec_1,\nvec_2,\ldots }
         V(\rvec_1+\nvec_1L,\rvec_2+\nvec_2L,\ldots ) \;.
\ee  
For example, the $x$-axis component of the two-body potential becomes
$$
\exp[-\Lambda^2 x_{ij}^2/4] \longrightarrow \sum_q \exp[-\Lambda^2 (x_{ij}-qL)^2/4],
$$
where in principle the sum over $q$ runs over all integers from minus infinity to
infinity. In practice, due to the short-range nature of the interaction, far boxes are
negligible and the sum is limited to $-N_{\mathrm{box}}\leq q \leq N_{\mathrm{box}}$. We have
verified that our results are fully converged for $N_{\mathrm{box}}=5$ 

%===========
\section{Methods}
\label{sec:methods}
%===========

To solve the $N$-body Schr\"odinger equation we first note that with the leading order
interactions \eqref{v2},\eqref{v3} spin and isospin are good quantum number. Thus we can
write the wave-function as a product of a spatial function times a spin state
$\chi_{S S_z}(\svec) $ and an isospin state $\xi_{T T_z}(\tvec) $ antisymmetrized to ensure 
Fermi statistics. Here $\svec=(s_1,s_2,\ldots s_A)$, and $\tvec=(t_1,t_2,\ldots t_A)$.
To satisfy the periodic boundary conditions we follow Ref. \cite{YinBlu13} and
expand the spatial part of the wave-function using a correlated Gaussians basis.
Using the abbreviations $\rvec=(\rvec_1,\rvec_2,\ldots \rvec_A)$ for the $A$-body
coordinates, and $\xvec = (x_1, \dots x_A)$ for the $x$ component (same for $y,z$) these basis 
functions are written as a product of periodic functions in the $x,y,z$ directions
\be
  G_L(\rvec)= G_{Lx}(\xvec)G_{Ly}(\yvec)G_{Lz}(\zvec) \;.
\ee
The $x$-component basis functions $G_{Lx}$ (same for $G_{Ly},G_{Lz}$) are
defined by a symmetric positive definite $A \times A$ matrices $A_x$, 
a positive definite diagonal matrix $B_x$, and a shift vector
$\dvec = (d_1, \dots d_N)$, 
\be 
G_{Lx} = \sum_{\nvec_x} G(A_x,B_x,\dvec_x;\xvec-L\nvec_x)
\ee
with
\be
 G = \exp\left[-\frac{1}{2}\xvec^TA_x\xvec-\frac{1}{2}(\xvec-\dvec)^TB_x(\xvec-\dvec)\right]
\;,
\ee
and ${\nvec_x} = (n_1, n_2, \dots n_A)$, $n_i \in \mathbb{Z}$.

The desired solution for the Schrodinger equation is given by
\be
  \Psi=\sum_k c_k \Phi(A_k,B_k,\dvec_k; \rvec, {\bf s}, {\bf t}),
\ee
where
\be
  \Phi_k = \hat{\mathcal A}\left[
    G_L(A_k,B_k,\dvec_k;\rvec) \chi_{S S_z} (\svec) 
    \xi_{T T_z} (\tvec)\right],
\ee
and $\hat{\mathcal A}$ is the antisymmetrization operator.
The linear parameters $c_k$ are obtained by solving the generalized eigenvalue problem
$H{\bs c}=EN{\bs c}$, 
where $H_{ij}=\bra \Phi_i | H | \Phi_j \ket $ are the Hamiltonian matrix elements,
and $N_{ij}=\bra \Phi_i | \Phi_j \ket$ the normalization matrix elements.
One of the advantages of the Gaussian basis is that the matrix elements
can be calculated analytically \cite{YinBlu13}.

To optimize our basis we use the Stochastic Variational Method (SVM) \cite{SuzVar98}.
To add a function to our basis, the elements of $A_k$, $B_k$, and $\dvec_k$ are chosen
randomly one by one, and the values which give the lowest energy are taken.

%==========
\section{Results}
\label{sec:results}
%==========
%~~~~~~~~~~~~~~~~~~~
\subsection{$m_\pi = 806$ MeV}
%~~~~~~~~~~~~~~~~~~~

First, we would like to deal with the results of the NPLQCD collaboration \cite{NPLQCD}.
These calculations assume $SU(3)$ symmetry, where the mass of the $u$ and $d$ quarks were
enlarged to the value of the $s$ quark mass. The resulting pion mass was calculated to be
$m_\pi = 806$ MeV and the nucleon mass was $m = 1634$ MeV.

Calculations were done for three lattice sizes, $L\approx 3.4$ fm, $4.5$ fm 
and $6.7$ fm.
The masses of light nuclei and hypernuclei with mass number $A\leq4$, and
strangeness $|s|\le 2$, were calculated. Here we focus on the nuclei, leaving 
hypernuclei for future publication.

Given a cutoff value, three data points are needed to calibrate the EFT. 
Here we choose to use the binding energies of the deuteron, di-neutron
(which is found to be bound for such heavy pion) and triton.

To verify that our results are cutoff independent we perform calculations with different
cutoff values (from 2~fm$^{-1}$ to 8 or 10 fm$^{-1}$). The results of the largest cutoff
values, which are fully converged, are shown hereafter. 

Tab. \ref{tbl:M806} summarizes the NPLQCD collaboration results \cite{NPLQCD} for 
the finite-volume binding energies of nuclei with $ A \le 4 $
calculated at pion mass of $m_\pi = 806$ MeV. 
three energies were calculated for each state,
corresponding to zero total momentum as well as to the two lowest boosted 
states.
For the largest lattice, we use all three states, while for the two smaller lattices 
the boosted states deviate from the ground state,
and therefore we did not use them. We checked, however, that our results do not change
substantially when all states are taken into account. 

\begin{table}
\begin{center}
  \caption{Light nuclei binding energies (in MeV) calculated in Ref. \cite{NPLQCD}
    using LQCD with $m_\pi = 806$ MeV for different lattice size $L$ (in fm).}
\label{tbl:M806}
\vspace{0.3cm}
{\renewcommand{\arraystretch}{1.25}%
\begin{tabular}
{c@{\hspace{5mm}} c@{\hspace{5mm}} c@{\hspace{5mm}} c}
\hline
\hline 
system & $L = 3.4$ & $L = 4.5$ & $L = 6.7$\\
\hline
nn          & $17.8\pm3.3$ & $15.1\pm2.8$ & $15.9\pm3.8$  \\
$^2$H   & $25.4\pm5.4$ & $22.5\pm3.5$ & $19.5\pm4.8$  \\
$^3$H   & $65.6\pm6.8$ & $63.2\pm8.0$ & $53.9\pm10.7$ \\
$^4$He & $ 115\pm 23$ & $107 \pm 25 $ & $107 \pm 25 $ \\
\hline
\hline
\end{tabular}}
\end{center}
\end{table}

Solving the Schr\"{o}dinger equation for each box size, we find the LECs that best fit the LQCD 
results of Tab. \ref{tbl:M806}, employing least-squares fit. The resulting LECs for several cutoff values 
are summarized in Tab. \ref{tbl:LECs}.

\begin{table}
\begin{center}
  \caption{The low energy constants of Eqs.\ref{v2},\ref{v3} fitted to the binding energies of Ref. \cite{NPLQCD}
    for $m_\pi = 806$ MeV, for different cutoff values $\Lambda$.}
\label{tbl:LECs}
\vspace{0.3cm}
{\renewcommand{\arraystretch}{1.25}%
\begin{tabular}
{c@{\hspace{5mm}} c@{\hspace{5mm}} c@{\hspace{5mm}} c@{\hspace{5mm}} c@{\hspace{5mm}} c}
\hline
\hline 
$\Lambda (\rm{fm}^{-1})$ & 2 & 4 & 6 & 8 & 10 \\
\hline
$C_0 (\rm{MeV\, fm^3})$ & -773 & -277 & -161 & -113 & -86.1 \\
$C_1 (\rm{MeV\, fm^3})$ & -30.1 & -5.33 & -2.13 & -1.13 & -0.71 \\
$D (\rm{MeV\, fm^6})$    &  1556 & 146   & 37.0  & 14.6  &  7.24 \\
\hline
\hline
\end{tabular}}
\end{center}
\end{table}
 
The continuum binding energies can now be predicted by solving the Schr\"{o}dinger 
equation in the limit of $L\longrightarrow \infty$.

\begin{figure}
  \begin{center}
    \includegraphics[width=8.6 cm]{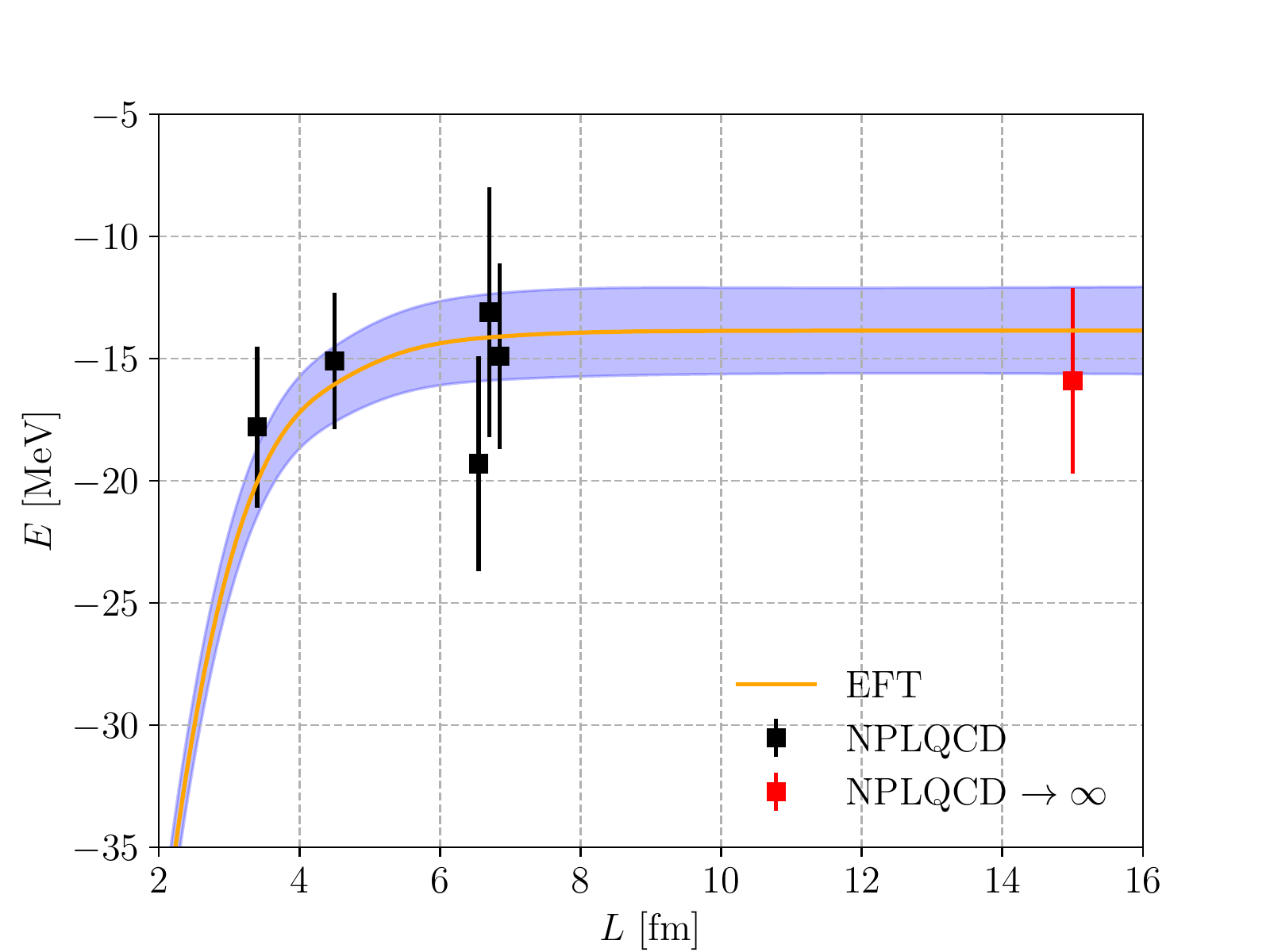}
    \caption{\label{fig:nn}
      The dineutron ground state energy as a function of the lattice size.
      EFT results (for $\Lambda=10$ fm$^{-1}$) are shown
      in blue curve, and the NPLQCD results used for fitting are shown in
      black squares. The red square shows Ref. \cite{NPLQCD} estimate for infinite lattice.}
\end{center}
\end{figure}

Two systems are bound in the nuclear two-body sector,
namely the deuteron and the di-neutron. 
The di-neutron binding energy calculated from the EFT is shown in
Fig. \ref{fig:nn} as function of lattice size.
The NPLQCD results, which were used to fit our EFT, are also shown.
The band stands for the error estimation of the EFT. The main source of the
error is the uncertainty and scatter of the LQCD results. 

\begin{figure}
  \begin{center}
    \includegraphics[width=8.6 cm]{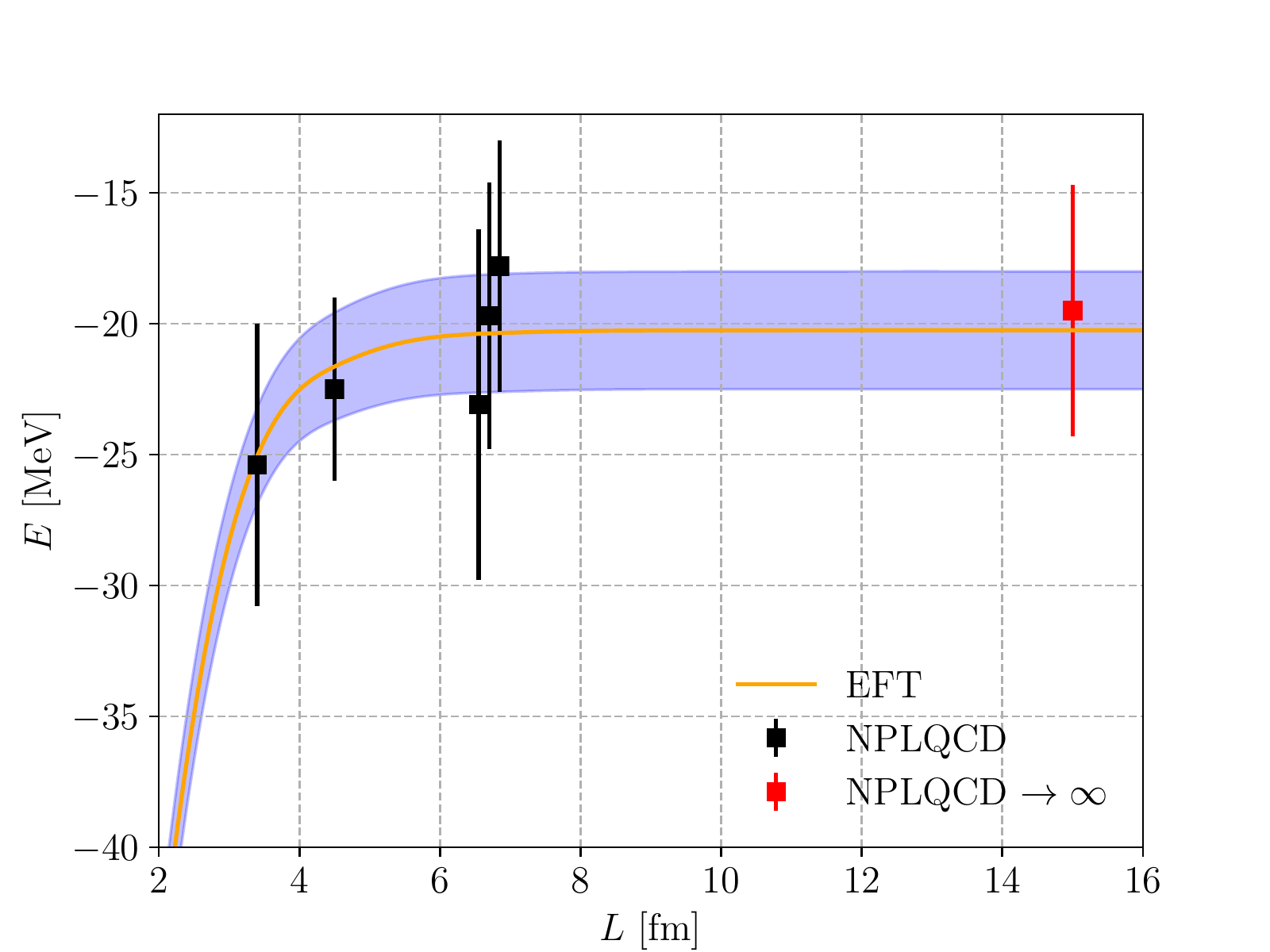}
    \caption{\label{fig:np}
      The deuteron ground state energy as a function of the lattice size.
      EFT results (for $\Lambda=10$ fm$^{-1}$) are shown
      in blue curve, and the NPLQCD results used for fitting are shown in
      black squares. The red square shows Ref. \cite{NPLQCD} estimate for infinite lattice.}
  \end{center}
\end{figure}

In Fig. \ref{fig:np} we show the deuteron binding energy calculated from the EFT, as well
as the data points from LQCD used for fitting. Also here the main source of error (shown as a
band) is the LQCD results. 

\begin{figure}
\begin{center}
\includegraphics[width=8.6 cm]{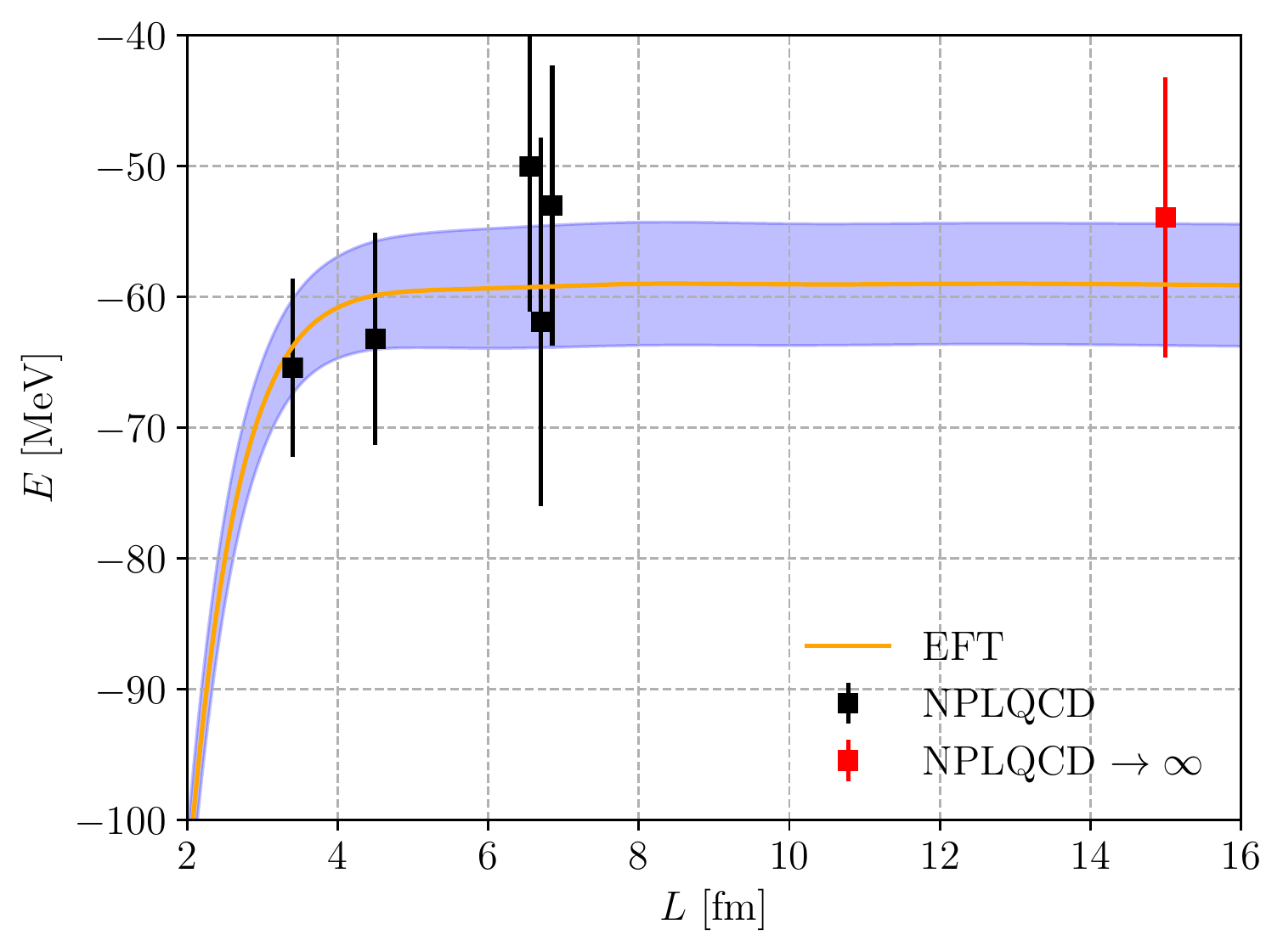}
\caption{\label{fig:He3}
  The triton ground state energy as a function of the lattice size.
  EFT results (for $\Lambda=10$ fm$^{-1}$) are shown
  in blue curve, and the NPLQCD results used for fitting are shown in
  black squares. The red square shows Ref. \cite{NPLQCD} estimate for infinite lattice.}
\end{center}
\end{figure}

Two bound states exist for nuclei in the three-body sector, namely $^3$H
and $^3$He. Following the LQCD calculations, we eliminate charge-symmetry breaking
terms as well as Coulomb forces and therefore their energies are degenerated. 
The triton ground state energy is shown in Fig. \ref{fig:He3} as a function of the lattice
size. Due to the deeper binding of the triton, its wavefunction is more compact and therefore
finite lattice corrections are less important, as one can see comparing
Fig. \ref{fig:He3} to Figs. \ref{fig:nn} and \ref{fig:np}.

\begin{figure}
\begin{center}
\includegraphics[width=8.6 cm]{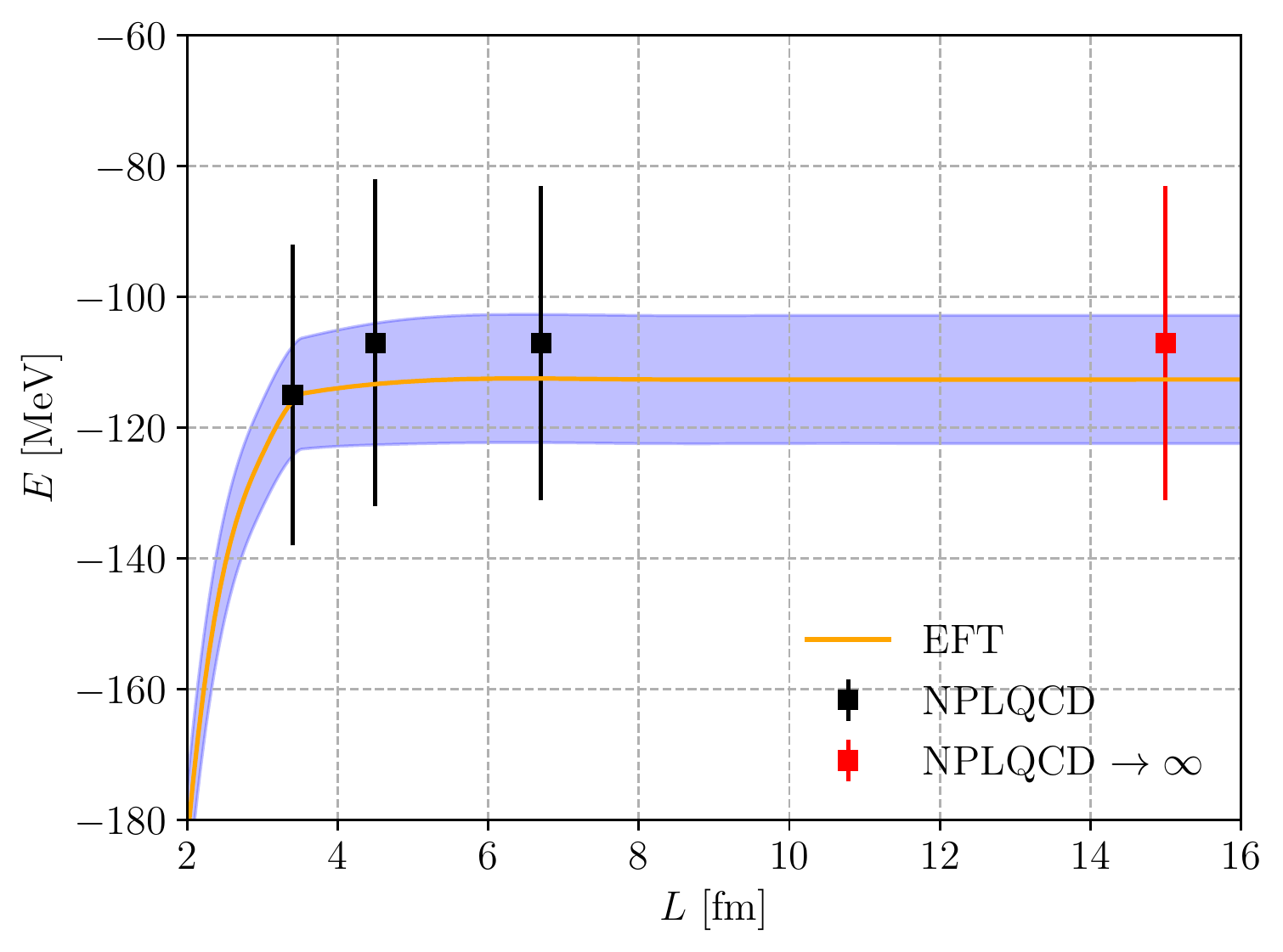}
\caption{\label{fig:He4}
  The $^4$He ground state energy as a function of the lattice size.
  EFT results (for $\Lambda=8$ fm$^{-1}$) are shown
  in blue curve, and the NPLQCD results used for fitting are shown in
  black squares. The red square shows Ref. \cite{NPLQCD} estimate for infinite lattice.}
\end{center}
\end{figure}

This effect is even more pronounced for the case of $^4$He, which is deeply bound for heavy 
pions, as can be seen in Fig. \ref{fig:He4}. 

Our extrapolated $L\longrightarrow\infty$ results are summarized in Tab. \ref{tbl:M806inf} and compared to the
values of the largest lattice which were taken as the infinite lattice limit in Ref. \cite{NPLQCD}. 
Our infinite volume results are consistent with the NPLQCD ones \cite{NPLQCD}, to within one standard deviation. 
The errors associated with our $L\to\infty$ extrapolations are smaller due to the use of more data points, associated 
with smaller error bars.
\begin{table}
\begin{center}
  \caption{Light nuclei binding energies (in MeV) from the largest 
    lattice of Ref. \cite{NPLQCD} and extrapolated to 
    to infinite lattice with out EFT.}
\label{tbl:M806inf}
\vspace{0.3cm}
{\renewcommand{\arraystretch}{1.25}%
\begin{tabular}
{c@{\hspace{5mm}} c@{\hspace{5mm}} c@{\hspace{5mm}} c}
\hline
\hline 
system & Ref. \cite{NPLQCD} & This work \\
\hline
nn          & $15.9\pm 3.8 $  & $13.8\pm1.8$ \\
$^2$H   & $19.5\pm 4.8 $  & $20.2\pm2.3$ \\
$^3$H   & $53.9\pm10.7$ & $58.2\pm4.7$ \\
$^4$He & $107 \pm 24  $ & $113 \pm10 $ \\
\hline
\hline
\end{tabular}}
\end{center}
\end{table}

%~~~~~~~~~~~~~~~~~~
\subsection{Physical pion mass}
%~~~~~~~~~~~~~~~~~~

In the near future, one would hope to see LQCD calculation for the physical
pion mass. Here we try to predict the lattice size corrections to the binding energies
of light nuclei, which may be utilized to choose the appropriate lattice size for such
calculations.

The dependence of the binding energy on the box size was studied in 
Refs. \cite{KorLuu16,DavSav11,BriDavLuu14} for the deuteron and 
in Ref. \cite{KorLuu16} for the triton. Here we compare our results for 
the deuteron and triton to those of Ref. \cite{KorLuu16} and make the first 
calculation, as far as we know, for $^4$He. 
Note, however, that the nuclear interaction used in this work, which is LO pionless EFT,
differs form the one used in Ref. \cite{KorLuu16}, LO $\chi $EFT. Both interactions has 
two-body contact terms; however, while $\chi $EFT has also one-pion exchange term, 
pionless EFT has additional three-body contact term. 

\begin{figure}
  \begin{center}
    \includegraphics[width=8.6 cm]{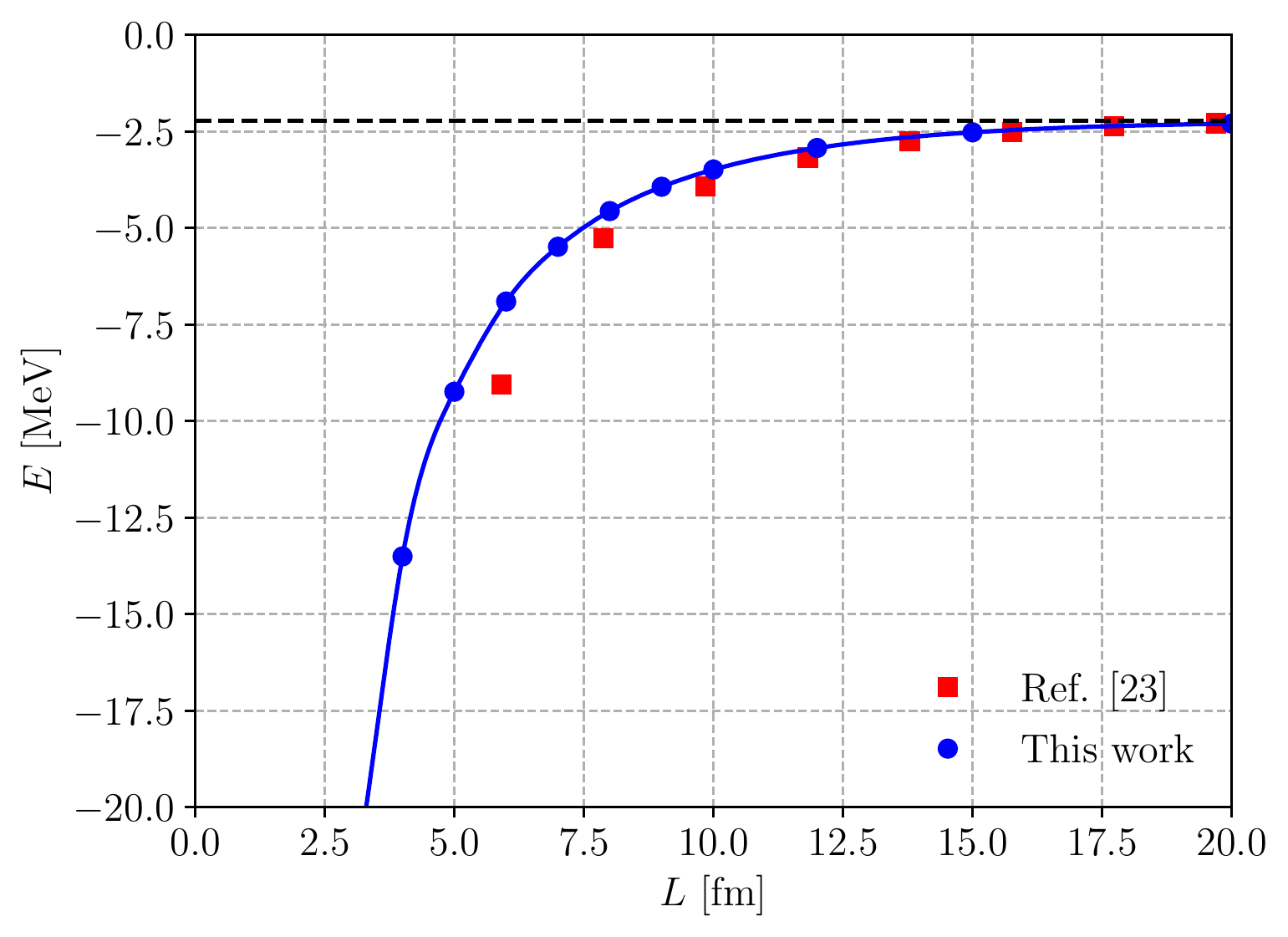}
    \caption{\label{fig:deuteron}
      The deuteron energy as a function of box-size for physical pion mass.}
  \end{center}
\end{figure}

The results for the deuteron are shown in Fig. \ref{fig:deuteron}. 
The deuteron binding is very shallow, resulting in a state with large spatial extent.
Consequently, the deuteron binding energy is converged to its asymptotic value 
only for very large lattices, $L \gtrsim 20$ fm.
This emphasizes the importance of extrapolation techniques for such calculations.
Comparing  our results to those of Ref. \cite{KorLuu16}, 
the results for large boxes seems identical, as one would expect since the asymptotic regime is 
govern by the binding momentum, see Eq. \ref{Luscher}.
Interestingly, for smaller boxes the results are different, due to the different potential used. 
This feature might be used to explore the short-range part of the nuclear interaction from 
LQCD calculation in finite lattices. 
 
\begin{figure}
  \begin{center}
    \includegraphics[width=8.6 cm]{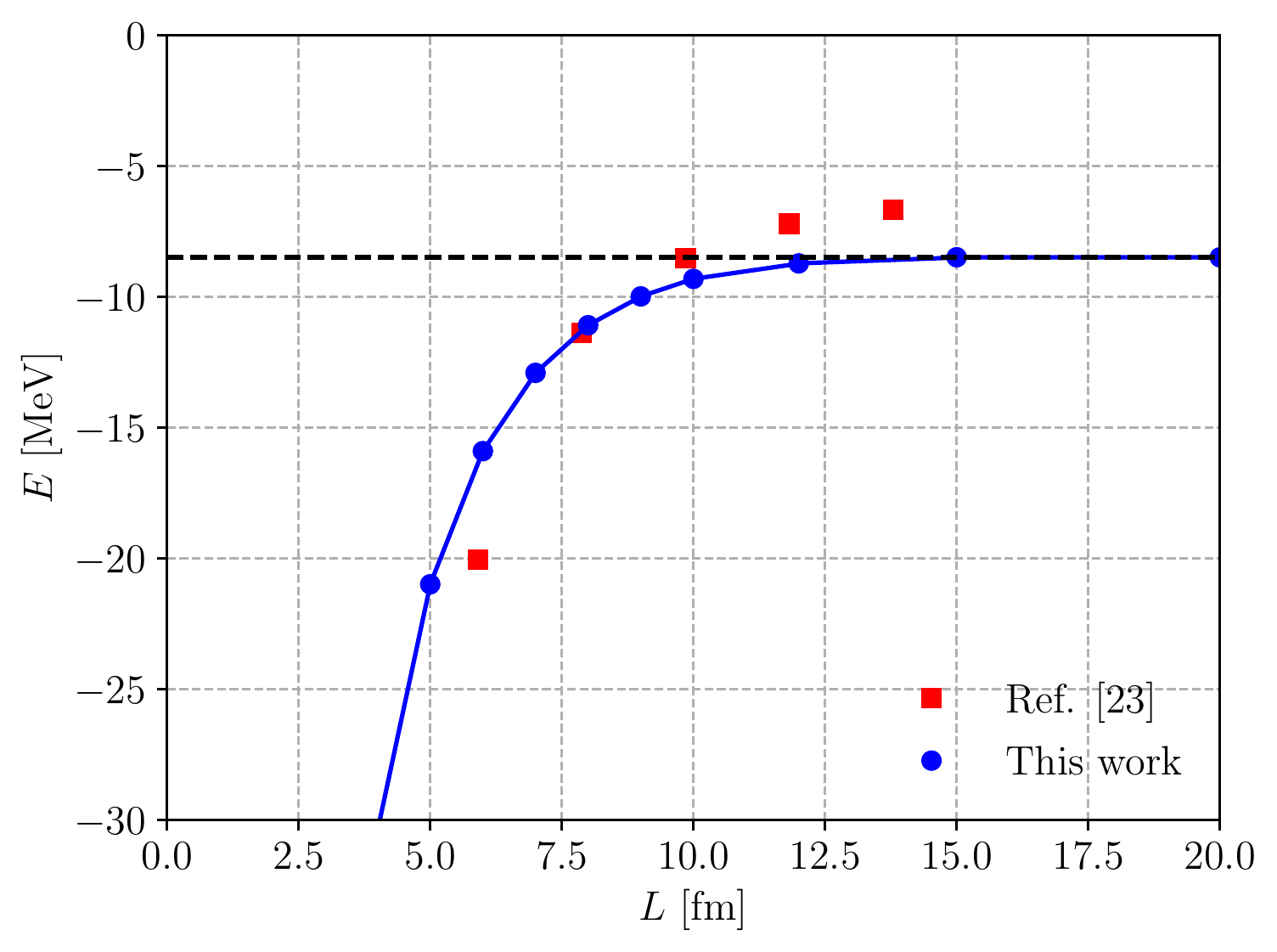}
    \caption{\label{fig:triton}
      The triton energy as a function of box-size for physical pion mass.}
  \end{center}
\end{figure}

The results for the triton are shown in Fig. \ref{fig:triton}. 
Here the results converge faster to the infinite volume limit, 
and the asymptotic value us retrieved at $L \gtrsim 12$ fm. 
It is harder to compare our results to those of Ref. \cite{KorLuu16} as 
they differ in the asymptotic values. This is evident since pionless
EFT has three-body term which is used to fit the triton binding energy, 
while $\chi $EFT has no additional degree of freedom at LO.

\begin{figure}
  \begin{center}
    \includegraphics[width=8.6 cm]{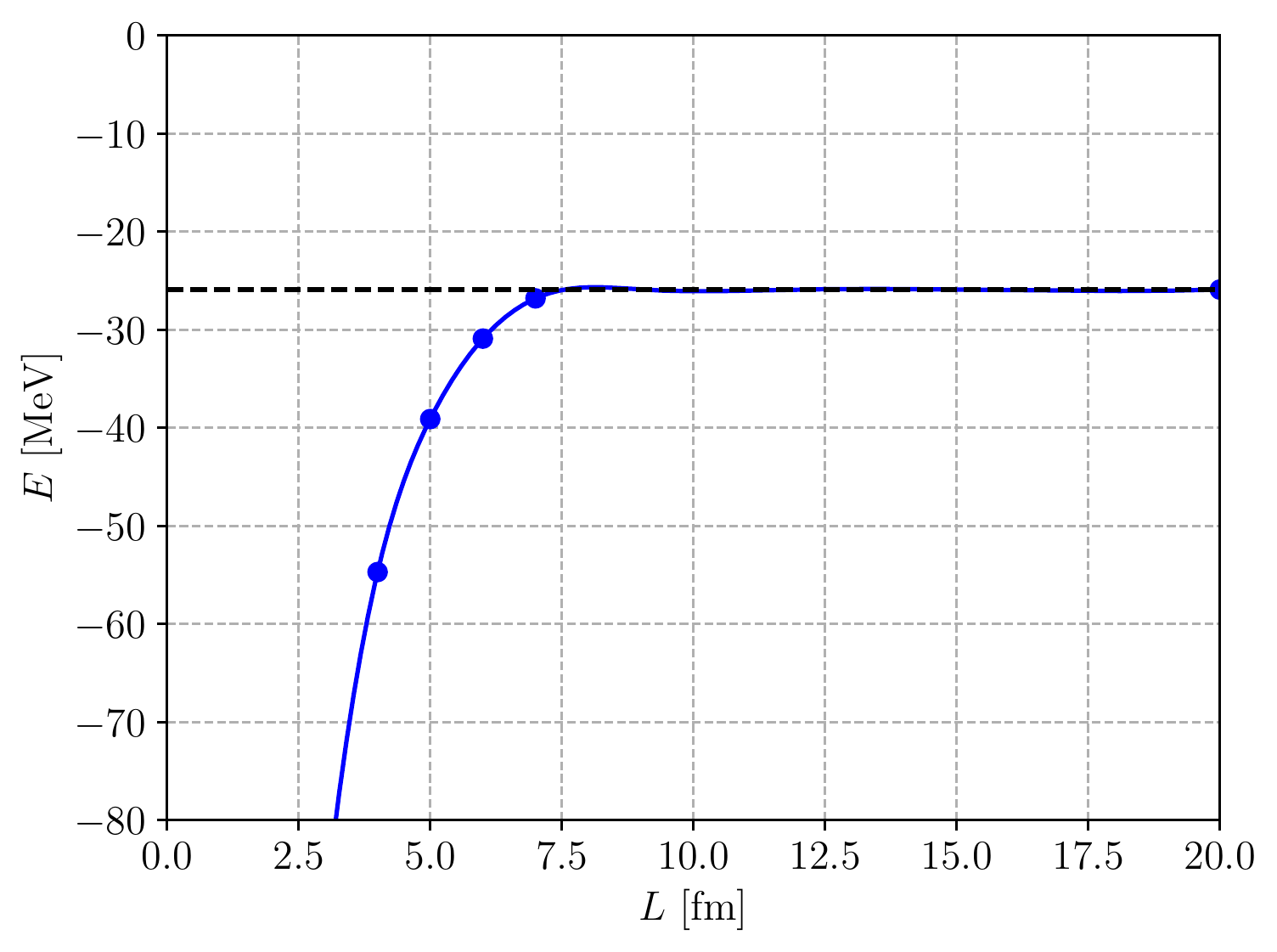}
    \caption{\label{fig:alpha}
      The $^4$He energy as a function of box-size for physical pion mass.}
  \end{center}
\end{figure}

Finally, In Fig. \ref{fig:alpha} the $^4$He binding energy is shown as a function of
the lattice size. Here the results converge even faster to the infinite volume limit,
and asymptotic results are obtained at $L \gtrsim 8$ fm. 

%===========
\section{Conclusion} 
%===========

The effect of the finite lattice size on the light nuclei binding energies is explored
by the construction of pionless effective field theory.
This theory, fitted to the LQCD results for small lattices, is then used to extrapolate
these results to the infinite volume limit.

We study the results of the NPLQCD collaboration for pion mass of $806$ MeV and present 
values for the infinite lattice limit. Our extrapolated binding energies are similar
to those extracted by the NPLQCD collaboration, albeit with smaller error bars
reflecting the use of more data points with better accuracy.

With an eye on future LQCD calculations at the physical pion mass, we predict the lattice size
correction for light nuclei, showing that he results are converged to the infinite lattice
size limit only at $L \gtrsim 20$ fm for the deuteron, at $L \gtrsim 12$ fm for the
triton and at $L \gtrsim 8$ fm for $^4$He. 
This emphasizes the importance of proper techniques to extrapolate the results
from small lattices.

%==============
\begin{acknowledgments}
%==============

  We would like to thank W. Detmold and L. Contessi
  for useful discussions during the preparation of this work,
  and T. Luu and C. K\"orber for providing us the data of Ref. \cite{KorLuu16}.
  This work was supported by the Israel Science Foundation (grant number 1308/16).
\end{acknowledgments}

%===============

\end{document}